\documentclass[prb,aps,twocolumn,showpacs,floats]{revtex4}
\usepackage{epsfig}

\begin{document}

\title{Density--functional study of Cu atoms, monolayers, and coadsorbates\\
       on polar ZnO surfaces}

\author{B. Meyer and D. Marx}
\affiliation{Lehrstuhl f\"ur Theoretische Chemie,
             Ruhr-Universit\"at Bochum, 44780 Bochum, Germany}
\date{\today}

\begin{abstract}
The structure and electronic properties of single Cu atoms, copper monolayers
and thin copper films on the polar oxygen and zinc terminated surfaces of ZnO
are studied using periodic density--functional calculations. We find that the
binding energy of Cu atoms sensitively depends on how charge neutrality of
the polar surfaces is achieved. Bonding is very strong if the surfaces are
stabilized by an electronic mechanism which leads to partially filled surface
bands. As soon as the surface bands are filled (either by partial Cu coverage,
by coadsorbates, or by the formation of defects), the binding energy decreases
significantly. In this case, values very similar to those found for nonpolar
surfaces and for copper on finite ZnO clusters are obtained. Possible
implications of these observations concerning the growth mode of copper on
polar ZnO surfaces and their importance in catalysis are discussed. 
\end{abstract}


\maketitle
\baselineskip11.4pt


\section{Introduction}
\label{sec:intro}

The ternary composite system Cu/ZnO/Al$_2$O$_3$ is an important commercial
catalysts for the methanol synthesis, the water--gas shift reaction and the
methanol--steam reforming process.\cite{cat1,cat2}  Although pure ZnO is a
good high temperature catalyst for hydrogenation and dehydrogenation
reactions, only the combination with Cu makes it an effective and selective
catalyst for low temperature applications on an industrial scale. In
particular the role of copper, including its electronic and atomistic
structure on ZnO surfaces, is not clearly understood. There exist many
controversial models to explain how copper enhances the activity of such
ZnO--based catalysts.\cite{rodriguez,campbell0}  Different ``active species''
are proposed in the literature, like the existence of Cu$^+$ impurities at
the ZnO surfaces, but the most commonly accepted picture is that Cu forms
small metallic clusters which are dispersed on defect--rich and heterogeneous
ZnO surfaces.

The initial growth of vapor--deposited copper on different ZnO surfaces at
ultra--high vacuum (UHV) conditions has been studied experimentally by
various techniques such as x--ray photoelectron spectroscopy
(XPS),\cite{campbell0,campbell1,campbell2,campbell3}
low--energy ion--scattering spectroscopy
(LEIS),\cite{campbell0,campbell1,campbell2,campbell3}
surface x--ray diffraction (SXRD),\cite{jedrecy} and scanning tunneling
microscopy (STM).\cite{diebold1}  For the two polar surface terminations of
ZnO a growth mode between a Volmer--Weber and Stranski--Krastonov behavior
was found. Initially, the deposited Cu atoms form two--dimensional islands
(i.e. clusters with a thickness of one atom) until a critical coverage
is reached, at which further Cu atoms begin to add on top of the islands
to form three--dimensional clusters, whereas the metal--free regions between
the clusters fill only very slowly.\cite{campbell1,campbell2}  This behavior
was attributed to {\em purely kinetic} limitations at low temperatures since
a thickening of the islands was observed upon annealing the surfaces. The
critical coverages at which three--dimensional growth of the clusters begins,
depends slightly on temperature and was estimated to be about 1/2 monolayer
for the O--terminated (000$\bar{1}$)--O surface\cite{campbell1} and 1/3 of a
monolayer for the Zn--terminated (0001)--Zn polar face of ZnO.\cite{campbell2}
In contrast, for clean, adsorbate--free nonpolar (10$\bar{1}$0) surfaces
exclusively three--dimensional islands were observed at all
coverages.\cite{diebold1}

The main focus of theoretical investigations of the bonding of metal
films and clusters on oxide surfaces has been on the MgO (001)
surface\cite{roesch1,musolino1,musolino2,musolino3,roesch2,roesch3}
and to a lesser extent on the surfaces of more complex oxides such as
TiO$_2$,\cite{tio2} Al$_2$O$_3$,\cite{finnis1,finnis2} and
MgAl$_2$O$_4$.\cite{koestlmeier}
Almost all studies were restricted to nonpolar surface terminations. Polar
surfaces, on the other hand, like the two basal planes of ZnO,
may show a completely different behavior compared to nonpolar surfaces. The
reason is that polar surfaces are characterized by an excess of one atomic
species\cite{finnis3} which leads to an interesting interplay between their
electronic and atomic structure: In order to be charge--neutral the polar
surfaces either have to adopt an electronic structure with partially occupied
surface bands (sometimes referred to as ``metallization'' of polar surfaces),
or they have to reconstruct (preferentially by the loss of suitable surface
atoms) in order to avoid partial band filling, or charged species have to be
adsorbed.\cite{nosker,noguera,zno37}  Depending on which mechanism is realized
in a particular system, very different properties of the interaction with
metal films and coadsorbates may occur.

Studies of the interaction of copper with ZnO surfaces are very scarce. In an
early semiempirical quantum--chemical study\cite{rodriguez} the electronic
structure and atomic charges were calculated for various ZnO clusters with
adsorbed and substituted Cu atom, but no information on binding energies and
relaxed atomic structures was obtained. In addition, clusters without
embedding were used which can not catch the peculiarities of the polar
surfaces described above. This was taken into account in a recent
quantum--chemical ab--initio study of the adsorption of single Cu atoms on
different cluster models\cite{fink} to which we will compare our periodic
density--functional results.

In the present study we have calculated the adhesion energy and geometry
of single Cu atoms and copper monolayers at various coverages on the two polar
surface terminations of ZnO. The surfaces are described by periodically
repeated slabs so that the different stabilization mechanisms for the polar
surfaces, in particular the existence of partially occupied surface bands, can
be handled without any restrictions. Additionally, atomic relaxations are
fully taken into account. The main focus of this study will be on how the
charge neutralization process of the polar surfaces and the filling of the
surface bands influences the adsorption properties of copper spanning the
range from single atoms to full monolayers. In addition to the very recent
quantum--chemical finite cluster study, Ref.~\onlinecite{fink}, these are the
first ab--initio results for the adhesion energies and geometries of copper on
ZnO surfaces.


\section{Computational Method}
\label{sec:theorie}

Self--consistent total--energy calculations based on density--functional
theory (DFT)\cite{hks} were carried out to determine the adsorption energies
and geometries of Cu atoms and monolayers on the polar ZnO surfaces.
The exchange--correlation energy and potential were treated within the
generalized--gradient approximation (GGA) using the functional of Perdew,
Burke and Ernzerhof (PBE).\cite{pbe}
Normconserving pseudopotentials\cite{van-pp} were employed together with a
mixed--basis set consisting of plane waves and non--overlapping localized
orbitals for the O--$2p$, the Zn--$3d$ and the Cu--$3d$ electrons.\cite{mb}
A plane--wave cut--off energy of 20\,Ry was sufficient to get well converged
results.

The polar surfaces are represented by periodically repeated slabs. No
mirror symmetry parallel to the polar surfaces is present which implies that
all slabs are inevitably Zn--terminated on one side and O--terminated on the
other. For the study of Cu monolayers (1$\times$1) surface unit cells and
very thick slabs consisting of 8 Zn--O double-layers were used to reduce the
residual internal electric field.\cite{zno34}. All atomic configurations
were fully relaxed by minimizing the atomic forces. The calculations for
single Cu atoms were performed on (2$\times$2) surface unit cells
(corresponding to a Cu surface coverage of 1/4 monolayer), and the slab
thickness was reduced to 4 ZnO double-layers. In a (2$\times$2) surface unit
cell arrangement the Cu atoms are more than 6.5\,{\AA} apart. The interaction
energy between the Cu atoms is calculated to be less than 0.02\,eV so that
this setup is sufficient to describe very well isolated Cu atoms.

\begin{figure}[!t]
\noindent
\epsfxsize=246pt
\epsffile{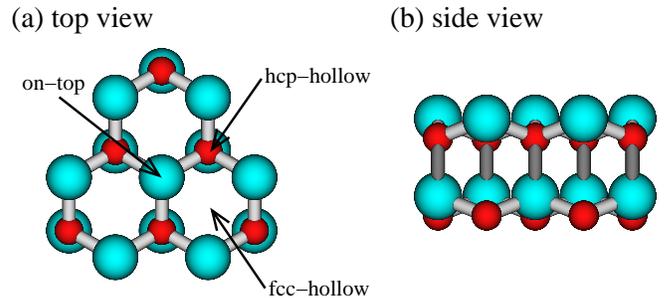}
\caption{\label{fig:polar}
Schematic diagram of the high--symmetry adsorption sites on the polar ZnO
surfaces, shown here for the oxygen surface termination only.}
\end{figure}

Three different high--symmetry adsorption sites are present at the polar
ZnO surfaces (see Fig.~\ref{fig:polar}): an `on-top' position which would
be the regular lattice site for the next atomic layer, a `hcp-hollow site'
above atoms in the second surface layer and a `fcc-hollow site' with no
atoms beneath. All three adsorption sites are considered in the present
study. Copper atoms are only adsorbed on one of the two slab terminations.
The electrostatic potential is calculated as for a genuinely isolated
slab\cite{bengtsson,bm} so no artificial interactions between the periodically
repeated images of the slabs is present. Monkhorst--Pack k--point
meshes\cite{mp} with a density of at least (6$\times$6$\times$4) points in the
Brillouin--zone of the primitive ZnO unit cell are chosen. For more details
on convergence parameters, the construction of appropriate supercells as well
as the calculated bulk and clean surface structures of ZnO we refer to
Ref.~\onlinecite{zno34}, where the same computational settings as in the
present study were used.

\section{Results and Discussion}
\label{sec:results}

\subsection{Validation of the Method: Finite Cluster Model}
\label{sec:val}

Before we start our investigation of copper on the polar ZnO surfaces, we
address the question how well our DFT method and in particular the PBE
functional are able to describe the interaction of Cu and ZnO surfaces. For
Cu on MgO(001) significant differences between various wave--function based
quantum--chemical methods and DFT calculations using different GGA functionals
were observed.\cite{roesch3}  As a general trend it was found that DFT
slightly overestimates the adhesion energy of Cu on MgO clusters compared to
best coupled--cluster calculations, with noticeable variations between the
GGA functionals.

\begin{figure}[!t]
\noindent
\epsfxsize=180pt
\centerline{\epsffile{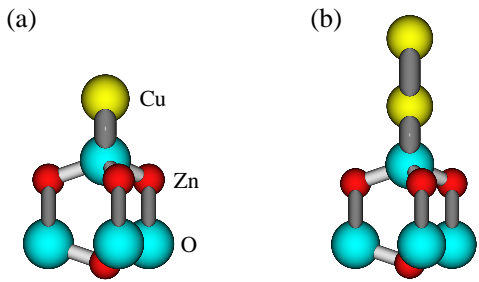}}
\caption{\label{fig:zn4o4}
Adsorption geometries of (a) single Cu atoms and (b) Cu dimers on-top of
a finite Zn$_4$O$_4$ cluster.}
\end{figure}

To test the reliability of our DFT results we have calculated the binding
energy and equilibrium distance of Cu on a small, isolated Zn$_4$O$_4$
cluster which we can compare with recently published results of an accurate
coupled--cluster--type calculation (multi--configuration coupled electron--pair
approximation MC--CEPA).\cite{fink}  A wurtzite--like cluster geometry was
chosen in order to mimic the stacking sequence of polar slabs (see
Fig.~\ref{fig:zn4o4}). The Cu atom is adsorbed to the top O and bottom Zn atom
of the cluster which are three--fold coordinated like at a ZnO surface.

The Zn$_4$O$_4$ cluster itself is a closed--shell system, but with the
additional Cu atom it becomes open-shell. Therefore, in the first calculation
for a single Cu atom on-top of the three--fold O atom of the cluster we used
the spin--polarized version of the PBE functional to obtain the total energy
of the adsorbate system as well as for the energy of the single Cu reference
atom. The binding energy and the equilibrium distance are given in
Table~\ref{tab:zn4o4}. No significant differences in the adsorption properties
are found if the spin--polarization is completely neglected and both, the
calculations for the adsorbate system and the separate Zn$_4$O$_4$ cluster and
the isolated Cu atom are done with the nonpolarized PBE functional. Also using
a Cu dimer instead of a single Cu atom gave a very similar result. Since the
inclusion of spin polarization doubles the computational cost, we only used a
Cu dimer for the study of the interaction of Cu with the three--fold
coordinated Zn atom of the finite cluster.

Compared to MC--CEPA results we find a satisfying agreement for the binding
energies and equilibrium distances. Like in the case of Cu on MgO, the DFT
calculation (here with the PBE functional) slightly overestimates the bond
strength. However, the adsorption energy is `equally' to large by 0.15 to
0.25\,eV, so that the relative stability of the O and the Zn adsorption site
is quite well described. Overall we find that Cu preferentially binds to
oxygen atoms with an adhesion energy of almost 1\,eV. The same behavior
was also found for MgO\cite{roesch3}, but with a lower binding energy of
0.4\,eV (coupled--cluster) and 0.6\,eV (DFT-GGA).

\begin{table}[!t]
\noindent
\begin{center}
\begin{minipage}[c]{246pt}
\def\arraystretch{1.2}
\def\tabcolsep{5pt}
\begin{tabular}{cc@{\quad}cc@{\quad}cc}
     & & \multicolumn{2}{c}{PBE} & \multicolumn{2}{c}{MC--CEPA}\\[4pt]
\parbox{46pt}{Adsorption\\Geometry} & & $d_0$ [\AA] & $E_{\rm ad}$ [eV]
                & $d_0$ [\AA] & $E_{\rm ad}$ [eV]\\ \hline
O--Cu      & sp & 1.91 & 0.98 & 1.95 & 0.73 \\
           & np & 1.91 & 0.96 \\
O--Cu$_2$  & np & 1.92 & 0.92 \\[8pt]
Zn--Cu     & sp &      &      & 2.80 & 0.12 \\
Zn--Cu$_2$ & np & 2.48 & 0.26 \\
\end{tabular}
\end{minipage}
\end{center}
\caption{\label{tab:zn4o4}
Equilibrium distances $d_0$ and binding energies $E_{\rm ad}$ of Cu atoms
and Cu$_2$ dimers on the Zn$_4$O$_4$ cluster for different adsorption
sites as indicated; `sp' refers to spin-polarized/open-shell and `np' to
nonpolarized DFT calculations. The MC--CEPA values are taken from
Ref.~\protect\onlinecite{fink}.}
\end{table}

\subsection{Cu Atoms on Ideal Polar Surfaces}
\label{sec:metallic}

For a long time it was believed that the polar ZnO surfaces exist in an
unreconstructed, truncated--bulk--like state.\cite{noguera,wander}
If no reconstruction by loss of surface atoms and no adsorption of
charged species occurs, the polar surfaces are inevitably stabilized
by partially filled surface states with a 3/4 and a 1/4 filled band for
the O and the Zn termination, respectively.\cite{nosker,noguera,zno37,wander}

Assuming such a situation in our DFT calculations we find a very strong
binding of Cu atoms on both polar ZnO surfaces (see Table~\ref{tab:metallic}).
The Cu binding energies are much larger than what is found for the 
finite Zn$_4$O$_4$ cluster (see Table~\ref{tab:zn4o4}) and, more importantly,
for the nonpolar ZnO(10$\bar{1}$0) surface\cite{zno39}.
In addition, it is also larger than what is typically known from other oxide
surfaces.\cite{roesch3}  The reason for this behavior is that at low copper
coverages, the $4s$ electrons of the Cu atoms fill the partially occupied
surface bands and thereby stabilize the polar surfaces. The same mechanism was
found for the adsorption of atomic hydrogen on the polar O--terminated surface
(see Ref.~\onlinecite{zno37}): at low H coverages the binding energy of an H
atom relative to a reference of isolated H$_2$ molecules is more than 2\,eV.
However, as soon as a H coverage of 1/2 monolayer is reached, the surface
bands are filled and the adsorption energy for additional H atoms decreases
rapidly. As a consequence, surface structures with more than 1/2 monolayer of
H are not stable in thermodynamic equilibrium with a H$_2$ gas phase at any
temperature and partial pressure.\cite{zno37}  Therefore, we can expect that
the strong binding of copper on the polar ZnO surfaces is only possible
as long as the surface bands are not fully occupied. If the bands are filled,
either by higher copper coverages or by the presence of coadsorbates like
hydrogen, the Cu adsorption properties may change drastically. However, for
copper there is one significant difference compared to hydrogen: at higher Cu
coverages strong Cu--Cu bonds will form so that Cu will also adsorb at
coverages exceeding 1/2 monolayer.

\begin{table}[!t]
\noindent
\begin{center}
\begin{minipage}[c]{246pt}
\def\arraystretch{1.5}
\def\tabcolsep{2pt}
\begin{tabular}{l@{\quad}cc@{\qquad}cc}
 & \multicolumn{2}{c}{(000$\bar{1}$)--O}
 & \multicolumn{2}{c}{(0001)--Zn}\\[4pt]
Adorption Site & $d_0$ [\AA] & $E_{\rm ad}$ [eV]
               & $d_0$ [\AA] & $E_{\rm ad}$ [eV]\\ \hline
on--top     & 1.84 & 2.57 & 2.45 & 1.81 \\
hcp--hollow & 1.93 & 1.85 & 2.08 & 1.87 \\
fcc--hollow & 1.25 & 2.87 & 1.97 & 2.06 \\
\end{tabular}
\end{minipage}
\end{center}
\caption{\label{tab:metallic}
Equilibrium distances $d_0$ and binding energies $E_{\rm ad}$ of single
Cu atoms for different adsorption sites on the ideal O-- and the
Zn--terminated polar ZnO surfaces with partially filled surface bands.
Slabs with a (2$\times$2) surface unit cell and a thickness of 4 Zn--O
double-layers were used for the calculation.}
\end{table}

\subsection{Cu Atoms on Adsorbate Pre--Covered Polar Surfaces}

Several very recent studies have created considerable doubt that the polar
ZnO surfaces really exist as ideal, truncated--bulk--like surfaces. With
scanning tunneling microscopy (STM) nanoscaled island and pit structures were
found at the Zn--terminated face\cite{diebold2} which lead to an overall
loss of roughly 1/4 of the surface Zn atoms. Such a stabilization mechanism
was confirmed in DFT calculations\cite{diebold2,kresse} which furthermore
predict high concentrations of adsorbed OH--groups in hydrogen--rich
environments. For the O--terminated surface it was shown by He atom
scattering,\cite{woell} in a study of the CO adsorption
energetics,\cite{zno35,zno36,staemmler} and also by DFT calculations
\cite{zno37} that they are H covered for a wide range of temperatures and
H partial pressures. Both mechanisms, loss of Zn atoms and adsorption of
hydrogen, lead to a filling of the surface bands. The filling need not to be
perfect, however, in order to study the consequences of such a stabilization
of the polar surfaces on the Cu (co)adsorption properties and also in order to
support our prediction from the last subsection, we have considered two
idealized surface structures of the polar ZnO surfaces for which the surface
bands are fully occupied.

For the O-- and the Zn--terminated surface we assume that 1/2 monolayer of
H atoms and OH groups, respectively, is present. We place the
H--atoms/OH--groups on every second atomic row at the surface. At the
adsorbate--free rows in between these adatoms the Cu atoms are placed and
the Cu adsorption energies are computed as in Sec.~\ref{sec:metallic}. 
The results are given in Table~\ref{tab:insulating}. We find indeed that
the Cu binding energy is significantly reduced compared to the result of the
ideal surface termination with partially occupied (``metallized'') surface
bands. The adsorption energy of roughly 1\,eV of Cu atoms on the H covered
O--terminated surface is only slightly larger than what was found for the
finite closed-shell Zn$_4$O$_4$ cluster and for the nonpolar ZnO surface
(having fully occupied bands, i.e. nonmetallic character).\cite{zno39} 
In addition, a similarly small value was also found in a quantum--chemical
embedded cluster study of the polar O--terminated surface\cite{fink} in which
the embedding scheme was chosen in such a way that it leads to fully occupied
surface bands, similar to our hydrogen covered surface.

\begin{table}[!t]
\noindent
\begin{center}
\begin{minipage}[c]{246pt}
\def\arraystretch{1.2}
\def\tabcolsep{5pt}
\begin{tabular}{lcc}
Adorption Site\quad\strut & $d_0$ [\AA] & $E_{\rm ad}$ [eV] \\ \hline
\multicolumn{3}{l}{(000$\bar{1}$)--O surface covered with
                   1/2 monolayer of H:}\\
\strut\qquad on--top     & 1.96 & 1.12 \\[4pt] \hline
\multicolumn{3}{l}{(0001)--Zn surface covered with 1/2 monolayer of OH:}\\
\strut\qquad fcc--hollow & 2.50 & 0.40 \\
\end{tabular}
\end{minipage}
\end{center}
\caption{\label{tab:insulating}
Equilibrium distances $d_0$ and binding energies $E_{\rm ad}$ of single
Cu atoms on polar ZnO surfaces with adsorbate structures to fill the
surface bands (which results in ``insulating surfaces'' for Cu coadsorption). 
Same slabs as in Table~\protect\ref{tab:metallic} were used.}
\end{table}

\subsection{Cu Monolayers on Polar Surfaces}
\label{sec:mono}

We turn now to the study of Cu monolayers on the polar ZnO surfaces. ZnO and
bulk Cu have very different lattice constants. The equilibrium distance of
the surface atoms at the polar surfaces is computed to be 3.28\,{\AA}
(experimentally: 3.25\,{\AA}), and for the Cu bulk lattice constant we found
3.65\,{\AA} which corresponds to a nearest--neighbor separation between the
Cu atoms of 2.58\,{\AA} (experimentally: 3.615\,{\AA} and 2.56\,{\AA},
respectively). Epitaxial growth with a (111) orientation of copper would
therefore lead to a lattice mismatch of 21\,\%. Thus, we can not expext that
expitaxial growth occurs. Experimentally it was found that Cu on the polar
surfaces is basically unstrained, even at very low coverages.\cite{jedrecy}
Exclusively (111) Cu planes were observed with a fixed orientation given by
the ZnO substrate.

The realistic treatment of such unstrained copper layers on the ZnO surfaces
would be computationally very demanding since we would have to use
prohibitively large surface unit cells for our slab. However, in order to get
a first idea on how the Cu adsorption strength and the preferred adsorption
sites change if a high concentration of copper is present on the polar
surfaces we have nevertheless studied ideal expitaxial Cu (111) monolayers on
the ZnO surfaces. For such hypothetical films we have calculated the
equilibrium distance and the work of separation which is defined as the energy
difference between the adsorbed monolayer and separate ZnO slabs and metal
films. The results are given in Table~\ref{tab:mono}. For the O--terminated
surface we find that Cu preferentially binds on-top of the oxygen atoms. The
equilibrium distance and the binding energy of 1.23\,eV per Cu atom are very
similar to the results obtained for single atoms on the finite Zn$_4$O$_4$
cluster and the surface with filled surface states. Also for the
Zn--terminated surface we find a good binding of the copper monolayers with
a work of separation of almost 0.9\,eV per Cu atom. This is in agreement with
experiment where a good adhesion of Cu on both polar surface terminations
was found.\cite{campbell1,campbell2}  In contrast to the O--terminated
surface, the hollow positions for the Cu atoms are more stable than the on-top
site. The two hollow positions are very similar in energy and may both be
populated for non--epitaxial films. This may explain the experimental
observation that the orientation of deposited Cu relative to the substrate
is less strict for the Zn--terminated face than for the
O--termination.\cite{campbell2,jedrecy}  Importantly, we also find that the
structural relaxation of the ZnO substrate can not be neglected. The
relaxation energy contributes 20\,\% to 30\,\% to the binding energy of these
copper films.

\begin{table}[!t]
\noindent
\begin{center}
\begin{minipage}[c]{246pt}
\def\arraystretch{1.2}
\def\tabcolsep{5pt}
\begin{tabular}{lcc@{\quad}cc}
 & \multicolumn{2}{c}{rigid slab} & \multicolumn{2}{c}{full relaxation}\\[4pt]
Adorption Site & $d_0$ [\AA] & $E_{\rm sep}$ [eV]
               & $d_0$ [\AA] & $E_{\rm sep}$ [eV]\\ \hline
(000$\bar{1}$)--O surface:\\
\strut\qquad on--top     & 1.98 & 1.02 & 1.94 & 1.23 \\
\strut\qquad hcp--hollow & 2.21 & 0.38 & 1.71 & 0.66 \\
\strut\qquad fcc--hollow & 1.75 & 0.58 & 1.36 & 0.90 \\ \hline
(0001)--Zn surface:\\
\strut\qquad on--top     & 2.56 & 0.54 & 2.50 & 0.62 \\
\strut\qquad hcp--hollow & 2.13 & 0.63 & 1.88 & 0.84 \\
\strut\qquad fcc--hollow & 2.12 & 0.66 & 1.89 & 0.87 \\
\end{tabular}
\end{minipage}
\end{center}
\caption{\label{tab:mono}
Equilibrium distances $d_0$ and work of separation $E_{\rm sep}$ (per surface
atom) of a Cu monolayer for different adsorption sites on the polar ZnO
surfaces, with and without atomic relaxation of the slab.}
\end{table}

\subsection{Cu Films on Polar Surfaces}
\label{sec:double}

Finally, we briefly study the effect higher copper coverages than full
monolayer on the bonding at the Cu/ZnO interface. This will give some insight
into the nature of the chemical bond at the interface. The higher coverages
were simulated by adding a second and a third monolayer of Cu on-top of the
polar surface. For the two and three layer films different stacking sequences
are possible. As shown in Table~\ref{tab:double} and Table~\ref{tab:triple},
the energy differences between different stacking sequences turn out to be
very small so that the specific stacking does not play a role in the following
arguments.

Overall we find that the same adsorption sites as for the monolayer coverage
are the most stable ones for the thicker Cu films: the on-top position for
the O--terminated surface and the hollow-sites for the Zn--terminated face.
For the O--terminated surface the work of separation increases with the
thickness of the copper film, whereas for the Zn--termination a small decrease
in the binding energy is observed. This behavior as well as the preferred
adsorption sites can be explained as follow: For the O--terminated surface
the first copper layer adopts the role of the missing next Zn layer. Therefore
the on-top position, which is the next regular lattice site of the wurtzite
structure, is preferred. Charge is donated to the surface O ions to fill the
partially occupied surface bands, which can more easily occur for thicker 
copper films, and a partially ionic/covalent bond like in bulk ZnO is formed.

On the other hand, for the Zn--terminated surface, the topmost Zn layer
behaves more like the first layer of the adsorbed metal film. Therefore
the hollow positions, which correspond to an extension of the metal film,
are the low-energy adsorption sites. The $4s$ electrons of the partially
filled Zn surface band contribute to the metallic state of the metal film
so that the Zn--Cu bond is more metallic in character. However, the
``integration'' of the topmost Zn layer becomes weaker for thicker copper
films.

\begin{table}[!t]
\noindent
\begin{center}
\begin{minipage}[c]{246pt}
\def\arraystretch{1.2}
\def\tabcolsep{5pt}
\begin{tabular}{lcc@{\quad}cc}
Adorption Site & sequence & $d_0$ [\AA] & $d_0$ [\AA] & $E_{\rm sep}$ [eV] \\
\hline
(000$\bar{1}$)--O surface:\\
\strut\qquad on--top & hcp & 1.96 & 1.53 & 1.50 \\
                     & fcc & 1.96 & 1.53 & 1.50 \\
(0001)--Zn surface:\\
\strut\qquad hcp--hollow & hcp & 1.86 & 1.48 & 0.94 \\
                         & fcc & 1.87 & 1.48 & 0.93 \\
\strut\qquad fcc--hollow & hcp & 1.90 & 1.47 & 0.91 \\
                         & fcc & 1.93 & 1.48 & 0.90 \\
\end{tabular}
\end{minipage}
\end{center}
\caption{\label{tab:double}
Equilibrium distances $d_0$ and work of separation $E_{\rm sep}$ (per surface
atom) of Cu double--layers for different adsorption sites on the polar ZnO
surfaces and different stacking sequences.}
\end{table}

\begin{table}[!t]
\noindent
\begin{center}
\begin{minipage}[c]{246pt}
\def\arraystretch{1.2}
\def\tabcolsep{5pt}
\begin{tabular}{lccc}
Adorption Site & sequence & $d_0$ [\AA] & $E_{\rm ad}$ [eV] \\ \hline
(000$\bar{1}$)--O surface:\\
\strut\qquad on--top & hcp & 1.96 & 1.33 \\
                     & fcc & 1.96 & 1.32 \\
(0001)--Zn surface:\\
\strut\qquad hcp--hollow & hcp & 1.87 & 0.76 \\
                         & fcc & 1.88 & 0.76 \\
\end{tabular}
\end{minipage}
\end{center}
\caption{\label{tab:triple}
Equilibrium distances $d_0$ and work of separation $E_{\rm sep}$ (per surface
atom) of Cu triple--layers on the polar ZnO surfaces for different stacking
sequences.}
\end{table}


\section{Summary and Conclusions}
\label{sec:summary}

%
Using DFT calculations and periodically repeated slabs to represent the two
polar ZnO surfaces, i.e. the (000$\bar{1}$)--O and (0001)--Zn faces, we find
a pronounced difference in the adsorption energy of Cu atoms depending on the
coverage. Crucial to the proposed understanding of this phenomenon is the fact
that the ideal, unreconstructed O-- and Zn--terminated surfaces feature
partially filled surface states with a 3/4 and a 1/4 filled band,
respectively (often referred to as ``metallization'' of the polar surfaces).
Up to 1/2 monolayer coverage the adhesion of copper on such polar surfaces is
found to be very strong since the $4s$ electrons of the Cu atoms can fill
these partially occupied bands and thereby stabilize the polar surface itself.
However, as soon as these bands are filled due to copper (or due to
coadsorbates!) already present on the surface, the binding of additional Cu
to the surface becomes weaker and less favorable in comparison to the cohesion
energy of Cu in small Cu clusters and thin Cu films.

Considering the experimentally known coverage--dependent change of the growth
mode of copper on the polar ZnO surfaces (i.e. formation of two--dimensional
islands until critical coverages of about 50\,\% and 30\,\% for the O-- and
the Zn--terminated surface, respectively, is reached at which
three--dimensional growth of Cu clusters sets in) it may very well be that
this sudden change is not caused by {\em kinetic} limitations at low
temperatures alone. Instead, it might also be driven {\em thermodynamically}
by the significant change in the adsorption energetics of copper on such polar
surfaces due to successive band filling, which in turn is directly coupled to
increasing the surface coverage. The lower critical coverage found
experimentally for the Zn--terminated surface may then be explained  quite
naturally by the lower adsorption energy of copper found for this particular
surface. Additionally, the presence of coadsorbates during or prior to copper
deposition may strongly influence both the critical coverage and the
morphology of the grown Cu particles.

%

Thus, it becomes evident that the {\em polar} ZnO surfaces allow for a wealth
of quite different adsorption scenarios depending on parameters such as
coverage, defects, coadsorbates etc. Since these parameters typically change
in complex ways during catalytic cycles one is tempted to speculate that the
polar character might be at the very heart of the catalytic activity of ZnO.


\section{Acknowledgments}

We are grateful to Karin Fink for making the results of the cluster
calculations available to us prior to publication, and we wish to thank
her and Volker Staemmler for fruitful discussions. The work was supported by
SFB~558 and FCI.



\end{document}